# High-Performance Nonvolatile Spin FETs from 2D Metallic Ferromagnetic and Ferroelectric Multiferroic Heterostructure

B. Liu[1], X. Zhang[2,‡], W. Hou[1], H. Feng[1], Z. Dai[1,‡], and Zhi-Xin Guo[1,†]

[1]*State Key Laboratory for Mechanical Behavior of Materials, Xi'an Jiaotong University, Xi'an, Shaanxi, 710049, China.*

[2]*Shaanxi Key Laboratory of Surface Engineering and Remanufacturing College of Mechanical and Materials Engineering Xi'an University, Xi'an 710065, China.*

†zhangxian1@outlook.com

‡sensdai@mail.xjtu.edu.cn

*zxguo08@xjtu.edu.cn

**Abstract:** All-electric-controlled nonvolatile spin field-effect transistors (SFETs) based on two-dimensional (2D) multiferroic van der Waals (vdW) heterostructures hold great promise for advanced spintronics applications. However, their performance is hindered by the limited availability of 2D magnetic materials that can switch effectively between metallic and semiconducting states with sizable bandgaps controlled by ferroelectric polarization. Most studies have focused on materials that are naturally semiconducting, achieving a metallic state by modifying the ferroelectric polarization. In this work, we introduce an innovative approach that uses interface effects to convert inherently metallic 2D magnetic materials into half-metals and induce half-semiconducting behavior through changes in ferroelectric polarization. Density functional theory (DFT) calculations on the $CrPS_3/Sc_2CO_2$ heterostructure demonstrate that the ferroelectric polarization of $Sc_2CO_2$ monolayers can adjust the electronic structure of $CrPS_3$, enabling a switch from half-metallic to half-semiconducting states. Building on these insights, we designed a nonvolatile SFET and analyzed its transport properties using the nonequilibrium Green's function (NEGF) method combined with DFT. Our results show that reversing the ferroelectric polarization achieves an on/off current ratio exceeding $5\times10^6$ %, and the heterostructure generates nearly 100 % spin-polarized current with a current density of up to 6500 µA/µm at bias voltage below 0.2 V. These findings highlight a promising pathway for developing high-performance SFETs that surpass existing 2D heterojunction materials.

# 1. Introduction:

Spintronics is believed to be a strong candidate of the future technology in the post-Si era. Practical applications of spintronics rely on key processes such as spin injection, detection, manipulation, and storage [1,2]. All-electric-controlled spin field effect transistor (SFET) is a fundamental spin-based device for the spin operation, which has attracted considerable research attention [3-6]. Advancing spintronics requires the development of SFETs that combine low power consumption, high performance, and efficient pure spin current switching.

A promising candidate for efficient SFETs is two-dimensional (2D) ferromagnetic (FM) half-metallic materials, which can generate 100% spin-polarized current by exhibiting metallic behavior in one spin channel and semiconducting or insulating behavior in the other. However, intrinsically half-metallic 2D materials are rare [7-10]. Although external modulation techniques such as electric fields and strain can induce half-metallicity [11], these approaches lack nonvolatility, limiting their practical applications. Recent progress in van der Waals (vdW) multiferroic heterojunctions, combining 2D ferroelectrics with 2D FM materials, offers a promising alternative. These multiferroic heterojunctions enable fully electric-controlled gating effects, paving the way for low-power and nonvolatile SFETs [12-21].

In the multiferroic heterojunctions, 2D ferroelectrics not only provide nonvolatility but also open new possibilities for spintronic devices. To date, research on 2D multiferroic heterojunctions has primarily focused on materials with intrinsic half-semiconducting properties, such as H-FeCl$_2$ [15], VSi$_2$N$_4$ [14,19], GdF$_2$ [21], and MnI$_3$ [20]. In these studies, metallic states are achieved by altering the ferroelectric polarization of the ferroelectric material. However, the performance of SFETs is limited by the scarcity of 2D semiconducting magnetic materials that can effectively switch between metallic and semiconducting states with sizable bandgaps controlled by ferroelectric polarization.

This study proposes a novel method that leverages interface effects to transform intrinsically metallic 2D magnetic materials into half-metals and induce half-semiconductor characteristics through changes in the ferroelectric polarization of 2D ferroelectric materials. As a proof of concept, we selected the metallic 2D magnetic material CrPS$_3$ and the 2D ferroelectric material Sc$_2$CO$_2$ to construct a vdW multiferroic heterojunction for use as the channel material in SFETs. First-principles calculations reveal that the monolayer Sc$_2$CO$_2$ effectively modifies the electronic structure of CrPS$_3$, enabling a transition from half-metallic to semiconducting states. The corresponding SFET demonstrates exceptional performance. Reversing the ferroelectric polarization of Sc$_2$CO$_2$ achieves an extraordinarily high *on/off* current ratio exceeding 5×10$^6$ % under finite bias. Moreover, the heterostructure generates nearly 100 % spin-polarized current with a current density of up to 6500 μA/μm at voltages below 0.2 V.

**Results and discussion:**

The heterostructure consists of a monolayer CrPS$_3$ and a monolayer Sc$_2$CO$_2$. As shown in Fig. 1(a), freestanding monolayer CrPS$_3$ is a magnetic material with a triangular lattice, where Cr atoms reside within octahedra formed by S atoms, and its optimized lattice constant is 5.90 Å, consistent with previous studies [22-24]. Although the freestanding monolayer CrPS$_3$ was reported to hold an antiferromagnetic (AFM) ground state [24,25], we find it has a FM ground state (180 meV more stable than the AFM state) in the heterostructure, as discussed below. As shown in Fig. 1(b), the FM phase of monolayer CrPS$_3$ exhibits excellent metallic characteristic, with a mix of spin-up and spin-down electronic bands around the Fermi energy ($E_F$). On the other hand, monolayer Sc$_2$CO$_2$ (Fig. 1(c)) also possesses a triangular lattice and is a ferroelectric material (out-of-plane polarization of 1.60 μC/cm²) with an optimized lattice constant of 3.43 Å, in agreement with prior works [26,27]. As depicted in Fig. 1(d), its band structure displays a large bandgap of approximately 1.84 eV between the VBM and CBM, consistent with previous studies.

In the heterostructure, a 1×1 unit cell of CrPS$_3$ is commensurate with $\sqrt{3} \times \sqrt{3}$ unit cell of Sc$_2$CO$_2$, resulting a lattice mismatch less than 5 %. As illustrated in Fig. S1, we explored four typical stacking configurations, with the corresponding energy calculations presented in Table S1. The lowest energy structure is shown in Fig. 2. The structural distortion at the CrPS$_3$-Sc$_2$CO$_2$ interface, induced by the presence of Sc$_2$CO$_2$, causes a downward shift of phosphorus atoms, which significantly alters the electronic structure. Notably, the ground state of CrPS$_3$ exhibits an AFM state [24,25], warranting an investigation into the magnetic ground state of the heterostructure. We calculated four magnetic states, as depicted in Fig. S2. The results indicate that regardless of the polarization direction, the CrPS$_3$ always has a FM ground state in the heterostructure, with the total energy being more than 180 meV lower than that of any other magnetic state (see Table S2).

It is noted that the interlayer surface atoms of the heterostructure are oxygen for Sc$_2$CO$_2$ and phosphorus atoms for CrPS$_3$, with atomic radii of 0.66 Å and 1.05 Å, respectively. Given that the interlayer distances in the P↑ and P↓ polarization states are 2.58 Å and 2.63 Å, which are substantially larger than the sum of the oxygen and phosphorus atomic radii, a vdW interaction exists between $CrPS_3$ and $Sc_2CO_2$. To ensure energetic stability, we further calculated the binding energy ($E_b$) between $CrPS_3$ and $Sc_2CO_2$, defined as $E_b=(E_{tot}-E_{CrPS}-E_{ScCO})/S$, where $E_{tot}$, $E_{CrPS}$, and $E_{ScCO}$ represent the total energy of the heterojunction, the energy of monolayer $CrPS_3$, and the energy of monolayer $Sc_2CO_2$, respectively, and S denotes the interface area of the simulation cell. In the P↑ and P↓ polarization states, $E_b$ values are -26.18 meV/Å² and -18.43 meV/Å², respectively, confirming the energetic stability of the heterostructure.

We now examine the influence of $Sc_2CO_2$ on the electronic structure of $CrPS_3$ which primarily governs the spin current. Despite the persistence of weak vdW interaction, we find that it still results in a sizable downward displacement of phosphorus atoms at the interface, significantly affecting the band structure of $CrPS_3$. Compared to the band structure of freestanding $CrPS_3$ [Fig. 1(b)], in the P↑ polarization state, the spin-down electronic band of $CrPS_3$ exhibits a substantial bandgap of 1.02 eV. Meanwhile, the spin-up electronic band intersects the Fermi level, transforming $CrPS_3$ into a half-metal with 100% spin polarization, as shown in Fig. 2(b). This confirms that the conversion of regular metallic $CrPS_3$ to a half-metal via interface effects has been successfully achieved. Furthermore, there are no electronic bands crossing the Fermi level, indicating minimal contribution of $Sc_2CO_2$ to the electronic transport in the heterostructure.

When the polarization of $Sc_2CO_2$ shifts to the P↓ state, the phosphorus atoms at the interface undergo a significant downward displacement due to the vdW interaction from $Sc_2CO_2$. Both the conduction band minimum (CBM) and the valence band maximum (VBM) in the spin-up and spin-down states are primarily contributed by $CrPS_3$, illustrating a type-I band alignment. This result demonstrates that the P↓ polarization of $Sc_2CO_2$ transforms $CrPS_3$ into a semiconductor with a large bandgap. Compared to the P↑ state where the $CrPS_3$ is a half-metal (zero/1.02 eV band gap for spin-up/spin-down electrons), in the P↓ state the $CrPS_3$ has a spin-up bandgap of 0.77 eV and a spin-down bandgap of 1.18 eV. This suggests that the electronic properties of $CrPS_3$ are heavily influenced by the out-of-plane polarization of $Sc_2CO_2$. It is noticed that in energy range around CBM, the electronic states are purely from spin-up electrons, making $CrPS_3$ a half-semiconductor characteristic. The above results manifest that reversing polarization of $Sc_2CO_2$ induces a half-metal to half-semiconductor transition in $CrPS_3$ with 100% spin-up polarization, which may have good performance in the

SFETs.

To further investigate the effects of $Sc_2CO_2$ polarization on electronic band structures of $CrPS_3$, we calculated the potential difference and charge transfer density of the heterostructure. Figures 3(a) and 3(c) illustrate the averaged potential differences for the heterostructure with spontaneous vertical polarization oriented upwards and downwards, respectively. This vertical polarization generates a potential difference, $\Delta\varphi$, between the upper and lower surfaces of the heterostructure. Reversal of the $Sc_2CO_2$ polarization direction changes the value of $\Delta\varphi$ from 1.17 eV in the P↑ state to 1.92 eV in the P↓ state. The difference in $\Delta\varphi$ indicates that once the electric field used to switch the polarization of $Sc_2CO_2$ is removed, the $Sc_2CO_2$ monolayer can maintain its polarization state, rendering the heterostructure nonvolatile. Significantly, the interface effects induce structural changes in $CrPS_3$, leading to pronounced ferroelectric polarization. The charge transfer density difference ($\Delta\rho$) for the heterostructure is described by the formula $\Delta\rho = \rho_{CrPS/ScCO} - \rho_{CrPS} - \rho_{ScCO}$, where $\rho_{CrPS/ScCO}$, $\rho_{CrPS}$, and $\rho_{ScCO}$ represent the charge densities of the $CrPS_3/Sc_2CO_2$ heterostructure, the monolayer $CrPS_3$, and the monolayer $Sc_2CO_2$, respectively (Figs. 3(b) and 3(d)). Integration of $\Delta\rho$ along the out-of-plane (z) direction reveals that in the P↑ state, electrons transfer from the surface of $Sc_2CO_2$ to $CrPS_3$, with a significant transfer of up to 1.5 electrons. Notably, the significant charge transfer from $Sc_2CO_2$ to $CrPS_3$ is mainly attributed to spin-up electrons, and the spin-down electrons primarily transfer to the interface center, as depicted in Fig. S3. On the other hand, when $Sc_2CO_2$ is polarized to the P↓ state, ferroelectric polarization induces minor charge transfer, i.e., the both electrons of $Sc_2CO_2$ and $CrPS_3$ transfer to the interface center (with only 0.47 electrons transferred). The polarization direction of $Sc_2CO_2$ strongly influences the charge transfer process [14], leading to distinct spin-polarized band structures in $CrPS_3$ (Fig. 2). This behavior is expected to enable distinguished nonvolatile SFET performance.

To verify the above prospection, we further designed a nonvolatile SFET device utilizing the vertically stacked heterostructure, as illustrated in Fig. 4(a). When $Sc_2CO_2$ is polarized to the P↑ state $CrPS_3$ generates a 100% spin-polarized current due to its half-metal characteristics. In this case, the SFET is in the "*on*" state. Conversely, when $Sc_2CO_2$ is polarized to the P↓ state, a significant bandgap in the $CrPS_3$ spin-up electronic band predicts a very small spin-up transmission current in the low bias voltage region, placing the SFET is in the "*off*" state. Note that the spin-down electronic band always exhibits a sizable bandgap in both P↑ and P↓ states, contributing to little transmission current. Hence, the heterojunction is expected to display rich nonvolatile spin transport characteristics with excellent device performance.

To validate these assertions, we calculated the electronic transport properties of the heterojunction SFET. Figures 4(b) and 4(c) depict the transmission spectra of the device under zero bias for $Sc_2CO_2$ in the P↑ and P↓ polarization states, respectively. In the P↑ state, in the low energy region of [0.05, 0.7] eV, the transmission rate of spin-up electrons is sizable with a maximum reaching 1.3 at 0.1 eV, while spin-down electrons exhibit nearly no transmission. As shown in Fig. 3(b) and Fig. S4, the sizable spin-up electron transmission is entirely contributed by $CrPS_3$ in such energy region, confirming the nature of 100% spin-polarized current in the SFET. When $Sc_2CO_2$ is in the P↓ state, the spin-down transmission spectrum remains zero across a substantial energy range of [-0.2, 0.8] eV, while the spin-up transmission spectrum is zero within the range of [-0.2, 0.2] eV. Thus, in the P↓ state, the spin current is fully suppressed within low finite bias. This result demonstrates that devices based on $CrPS_3/Sc_2CO_2$ possess excellent spin valve performance.

We additionally investigated the electronic transport behavior of the $CrPS_3/Sc_2CO_2$ device under low bias voltage of [0, 0.5] V. Figure 5(a) illustrates the total transmission current (the sum of spin-up and spin-down currents) and the corresponding *"on/off"* ratio. When $Sc_2CO_2$ is in the P↑ state, a large current is observed within the bias range 0-0.5 V. The current peak reaches at 6623 μA/μm appearing at a bias of 0.1 V and gradually decreases to 3400 μA/μm as the bias increases to 0.5 V, presenting a characteristic of negative differential conductance (NDC). To elucidate the underlying mechanism of the NDC, we further analyzed the transmission spectrum of different bias. As shown in Fig. S5, the transmission clearly decreases as the bias voltage increases from 0.1 to 0.5 V, leading to a significant decrease of the transmission current. On the other hand, in the P↓ state the current remains below 0.3 μA/μm within the bias voltage smaller than 0.5 V and decreases to 0.06 μA/μm at 0.5 V, showing extremely low leakage current. This feature results in remarkably large *"on/off"* ratio, greater than $2\times 10^6$ as shown in Fig. 5(a). Especially, the *"on/off"* reaches up to $5.4\times 10^6$ when the bias voltage reaches 0.5 V. In Fig. 5(b), we additionally show the ratio between spin-up/spin-down current for both P↑ and P↓ states. It is seen that the ratio also monotonously increases from $10^3$ ($10^4$) to $8\times 10^5$ ($6\times 10^9$) with the bias voltage increasing from 0.1 to 0.5 V for the P↑ (P↓) state, verifying that a nearly 100% spin-polarized current is achieved in the SFET device. These results manifest the excellent performance of $CrPS_3/Sc_2CO_2$ all-electric-controlled nonvolatile SFETs.

Finally, we conducted a comparative analysis of the $CrPS_3/Sc_2CO_2$ SFET device with other reported SFETs based on 2D multiferroic heterostructure reported so far. Table 1 summarizes a comparison on the three significant device parameters among our $CrPS_3/Sc_2CO_2$ SFET and other SFETs, i.e., *on/off* ratio, spin-up/spin-down current ratio and tunneling electroresistance (TER) (define as TER=$\frac{|T\uparrow - T\downarrow|}{\min(T\uparrow, T\downarrow)}$) [16]. It is seen that all these parameters are much more superior to other SFETs reported so far [14-16]. Moreover, the "*on/off*" ratio of the $CrPS_3/Sc_2CO_2$ SFET device is comparable to that of other conventional 2D vdW ferroelectric field-effect transistors (FeFETs), as shown in Table 1, highlighting the great potential of $CrPS_3/Sc_2CO_2$ for high-performance non-volatile SFET applications.

## Summary:

In summary, we discover that metallic 2D magnetic materials can be an ideal candidate in the multiferroic heterostructure for the high-performance nonvolatile SFET devices. By performing first-principles calculations on $CrPS_3/Sc_2CO_2$ heterostructure, we confirm that the interface effects from ferroelectric $Sc_2CO_2$ can convert inherently metallic magnetic $CrPS_3$ into half-metals and further induce half-semiconducting behavior through changes in ferroelectric polarization. Building on these insights, we designed a nonvolatile SFET based on the $CrPS_3/Sc_2CO_2$ heterostructure. Electronic transport calculations based on the ENGF method, further show that this SFET exhibits exceptional performance under low bias voltage region [0, 0.5] V. Our results show that reversing the ferroelectric polarization achieves an "*on/off*" current ratio in range of $2\times10^6$ % to $5.5\times10^6$ %. Moreover, the on-state currents are very large (3400-6500 μA/μm), presenting nearly 100% spin polarization where the ratio between spin-up/spin-down current is greater than $10^3$. The performance of $CrPS_3/Sc_2CO_2$ device significantly exceeds that of other nonvolatile SFET devices, manifesting the great potential applications of multiferroic heterostructure based on 2D metallic magnetic materials in the high-performance, low-power, non-volatile SFET devices.

## Method:

The density functional theory (DFT) with the projector-augmented-wave (PAW) method, which is implemented in the Vienna Ab Initio Simulation Package [28-30], are used for Geometric optimization and electronic structures of the heterostructure. The convergence standards of the atomic energy and positions are less than $1\times10^{-5}$ eV per atom and $1\times10^{-2}$ eV/Å, respectively. The cutoff energy of wave function is set to 520 eV. Uniform k meshes of 12 × 12 × 1 and 21 × 21 × 1 within the Γ-centered Monkhorst-Pack scheme [31] are adopted for the structural optimization and the self-consistent calculations, respectively. The exchange-correlation interaction is treated by the generalized gradient approximation (GGA) based on the Perdew-Burke-Ernzerhof (PBE) function [32] for the geometric optimization and electronic structures. The correlation effects for the d electrons of chromium are incorporated using the GGA+U [33] approach with an on-site effective interaction parameter U = 3.0 eV [34]. A vacuum layer of more than 15 Å is used, and the interlayer vdW interactions are treated with the DFT-D2 functional [35].

The transport properties are simulated based on the DFT method combined with the nonequilibrium Green's function (NEGF) formalism [36], via the Quantum-ATK package [37]. The K-point of device is set 7×1×93 in the transport calculations and the density mesh cutoff is set of 85.0 Hartree. The transmission calculations were carried out using the HGH Tier3 of the PBE pseudopotentials distributed in the Quantum-Wise package. The spin-resolved current at a bias voltage $V_b$ can be calculated as

$$I_\sigma(V_b) = \frac{e}{h}\int_{-\infty}^{+\infty} dE\,[f(E,\mu_L) - f(E,\mu_R)]\,T_\sigma(E,V_b). \tag{1}$$

Here the $f(E,\mu_L)$ and $f(E, \mu_R)$ the Fermi-Dirac distribution of the left and right electrodes, respectively. $\mu_L$ and $\mu_R$ are the electrochemical potentials of the left and right electrodes, respectively. The $T_\sigma(E, V_b)$ is the transmission probability for an electron at energy E with spin σ at a bias $V_b$.

## Acknowledgments

This work was supported by the Ministry of Science and Technology of the People's Republic of China (Grant No. 2022YFA1402901), Natural Science Foundation of China (No. 12474237,

## Conflict of interest

The authors declare no conflict of interest.

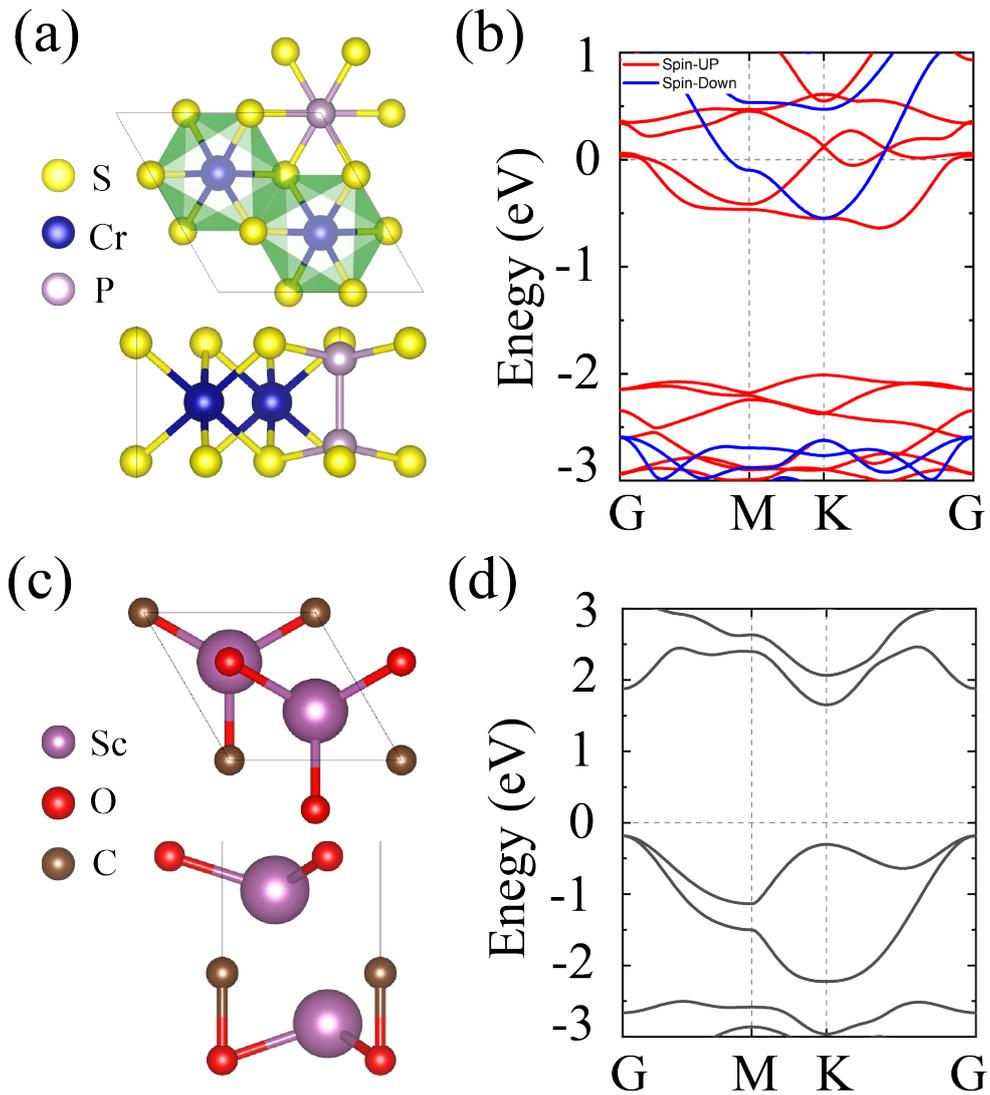

Figure 1 (a) Atomic structures of top view (top) and side view (bottom) of monolayer $CrPS_3$. (b) Band structure of FM of monolayer $CrPS_3$. (c) Atomic structures of top view (top) and side view (bottom) of monolayer $Sc_2CO_2$. (d) Band structure of monolayer $Sc_2CO_2$. The yellow, blue, light purple, great purple, red and brown represent S, Cr, P, Sc, O and C atoms, respectively.

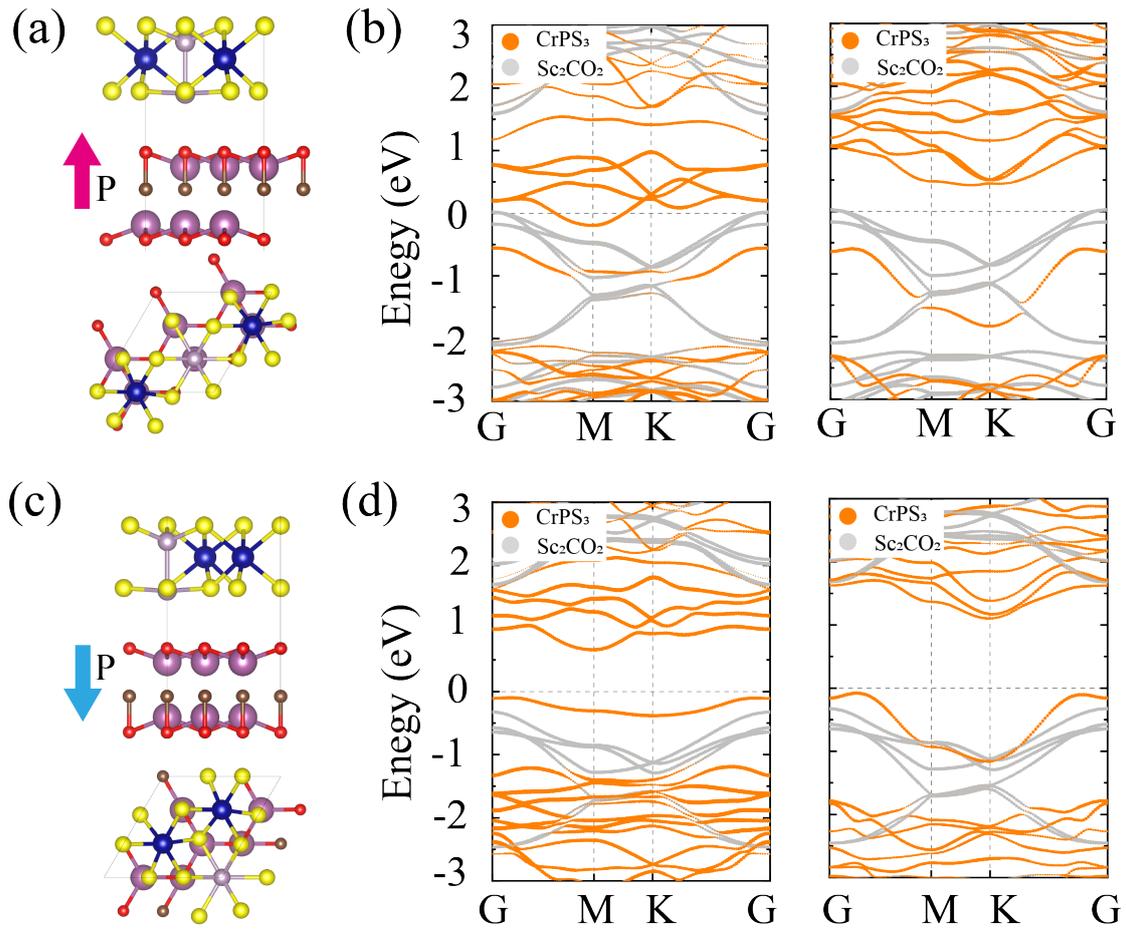

Figure 2 (a) and (c): Atomic structures of the $CrPS_3/Sc_2CO_2$ heterostructure with $Sc_2CO_2$ in (a) P↑ and (c) P↓ polarizations. (b) and (d): The electronic band structures of the $CrPS_3/Sc_2CO_2$ heterostructure with $Sc_2CO_2$ in (b) P↑ and (d) P↓ polarizations, where the left and right panels represent the spin-up and spin-down energy bands, respectively.

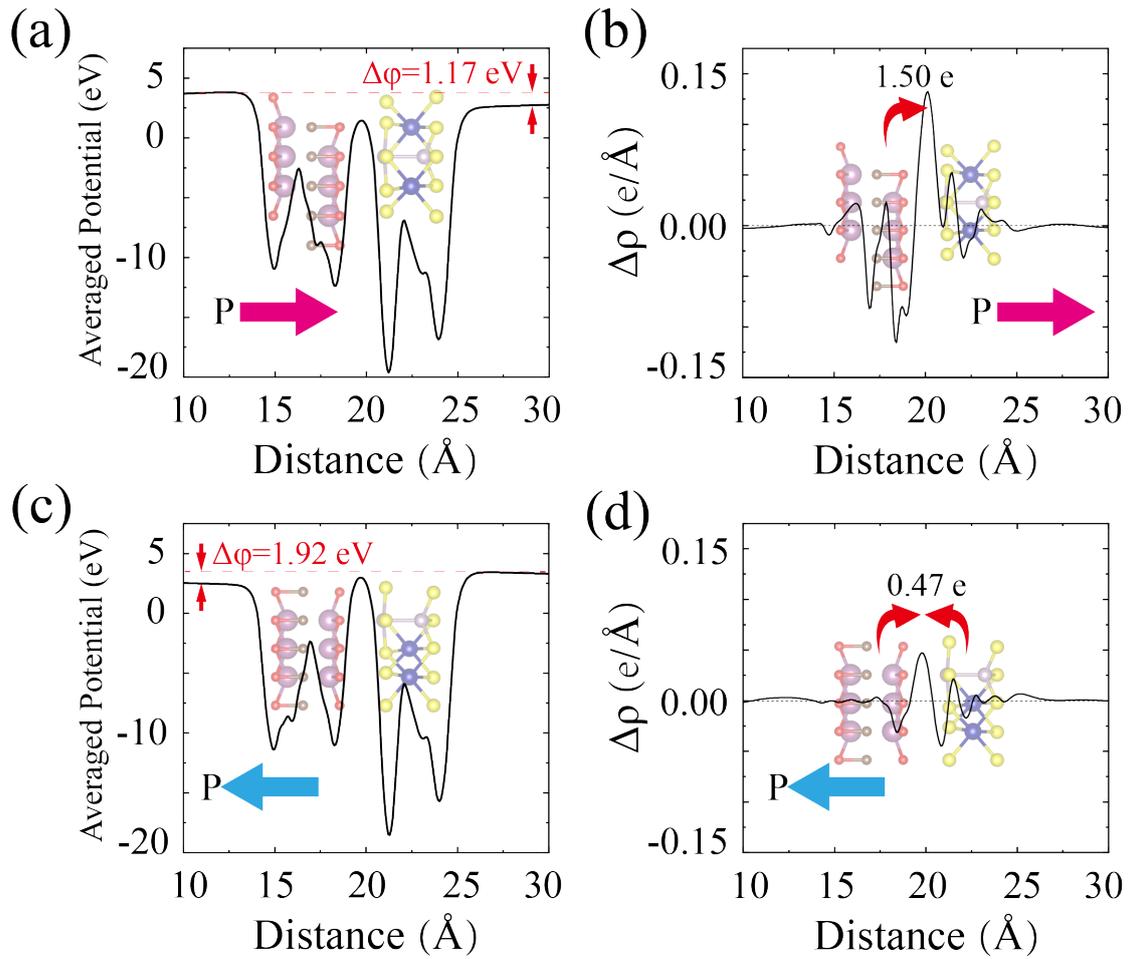

Figure 3 (a) and (c): Plane-averaged electrostatic potential of $CrPS_3/Sc_2CO_2$ (a) P↑ and (c) P↓ along the out-of-plane (z) direction. (b) and (d): The plane-averaged differential charge density Δρ of $CrPS_3/Sc_2CO_2$ (b) P↑ and (d) P↓ along the z direction.

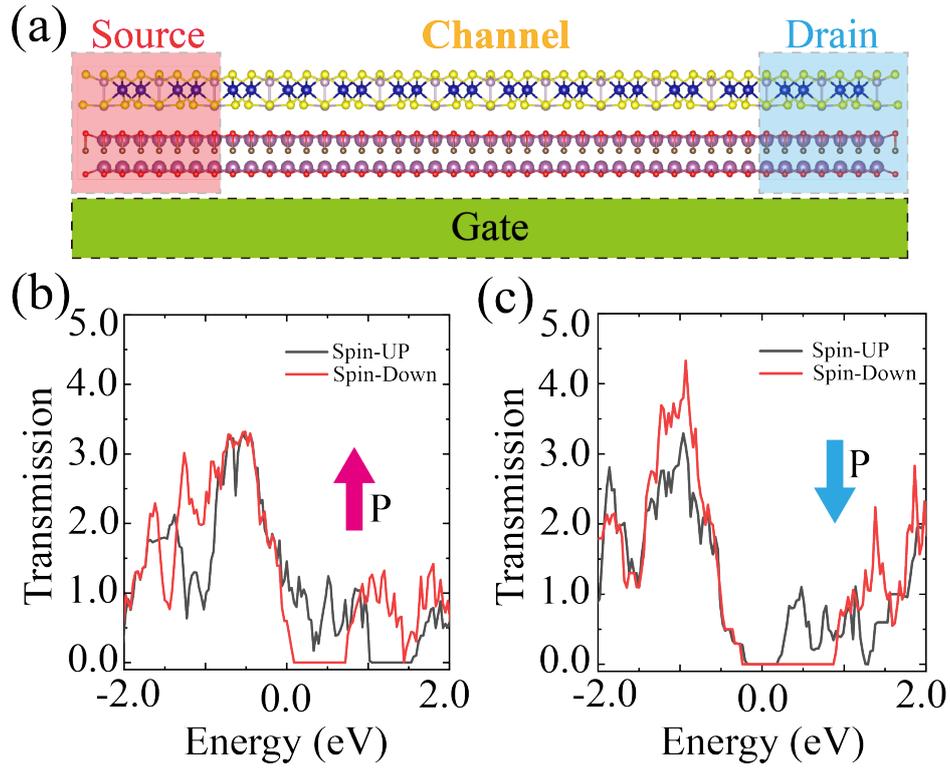

Figure 4 (a) The schematic structure of the proposed SFET based on the $CrPS_3/Sc_2CO_2$ heterostructure. (b) The calculated spin-polarized transmission spectrum of the SFET with $Sc_2CO_2$ in P↑ polarization at zero bias. (c) The calculated spin-polarized transmission spectrum of the SFET with $Sc_2CO_2$ in P↓ polarization at zero bias.

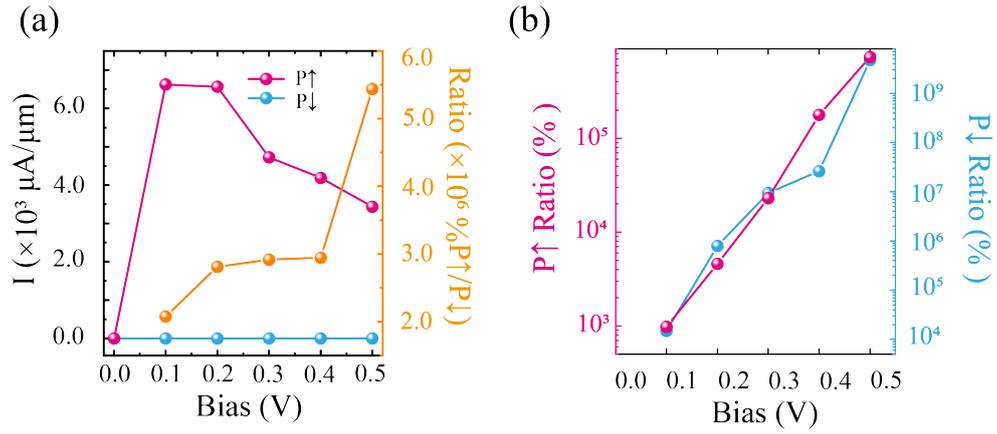

Figure 5 (a) The total current (sum of spin-up and spin-down current) with $Sc_2CO_2$ in P↑ (magenta line), P↓ (sky blue line) polarizations and *on/off* ratio (orange Line) under finite bias of 0.0-0.5 V. (b) The ratio of spin up and spin down current with $Sc_2CO_2$ in P↑ (magenta line), P↓ (sky blue line) polarizations under finite bias of 0.0-0.5 V.

Table 1 The *on/off* ratio, current of spin-up/spin-down ratio and TER of SFETs and FeFETs device.

| Materials | *on/off* ratio (%) | spin-up/spin-down ratio (%) | TER (%) |
|---|---|---|---|
| $CrPS_3/Sc_2CO_2$ | $5.4 \times 10^6$ | 730,000 | $1.6 \times 10^{23}$ |
| $FCl_2/Sc_2CO_2$ [15] | $3.2 \times 10^5$ | 155.12 | $1.7 \times 10^{17}$ |
| $VSi_2N_4/Sc_2CO_2$ [14] | $6.5 \times 10^2$ | 2,500 | $1.6 \times 10^{10}$ |
| $RuCl_2/Al_2S_3$ [16] | \ | 1,027 | $3.0 \times 10^7$ |
| $Cu/CuInP_2S_6(CIPS)/graphene$ [38] | $10^6$ | \ | \ |
| In/CIPS/hBN/InSe [39] | $10^6$ | \ | \ |
| $Au/CIPS/MoS_2$ [40] | $10^4$ | \ | \ |